\documentclass{nature}
\usepackage{graphicx}
\usepackage{amsmath}
\usepackage{amssymb}

\title{Fate of the Josephson effect in thin-film superconductors}

\author{Michael Hermele$^1$, Gil Refael$^2$, Matthew P. A. Fisher$^2$, Paul M. Goldbart$^3$}

\begin{document}

\maketitle

\begin{affiliations}
 \item Department of Physics, University of California Santa Barbara, Santa Barbara, California 93106, USA
  \item Kavli Institute for Theoretical Physics, University of California Santa Barbara, Santa Barbara, California 93106, USA
  \item Department of Physics, University of Illinois at Urbana-Champaign, Urbana, Illinois  61801, USA
\end{affiliations}

\begin{abstract}
The dc Josephson effect refers to the dissipationless electrical current -- the supercurrent -- that can be sustained
across a weak link connecting two bulk superconductors.\cite{josephson62}  This effect is a probe of the fundamental nature
of the superconducting state, which depends crucially on the spatial dimensionality of the superconducting electrodes.  For bulk (\emph{i.e.} three-dimensional) superconductors, the superconductivity is most robust, and the Josephson effect is sustained even at nonzero temperature.  However, in wires and thin-films (\emph{i.e.} lower-dimensional superconductors), thermal and quantum fluctuations play a crucial role.  In superconducting wires, these effects qualitatively modify the electrical transport across a weak link.\cite{kane92}  Despite a number of recent experiments involving weak links between thin-film superconductors,\cite{bezryadin00,lau01,naaman01, chu04} little theoretical attention has been paid to the nature of the electrical conduction in such systems.  
Here, we analyze the case of two superconducting thin films connected by a point contact.
Remarkably, the Josephson effect is absent at nonzero temperature, and 
the resistance across the contact is nonzero.
Moreover, 
the point contact resistance is found to vary with temperature in a nearly activated fashion,
with a \emph{universal} energy
barrier determined only by the superfluid stiffness 
characterizing the films,
an angle characterizing the geometry, and whether or not the
Coulomb interaction between Cooper pairs is screened.
This behavior reflects the subtle nature of the superconductivity in
two-dimensional thin films, and should be testable in detail by future experiments.   
\end{abstract}

Soon after the development of the microscopic theory of superconductivity,\cite{bcs} Josephson predicted
a remarkable manifestation of it: charge is transferred via a supercurrent across a weak link (\emph{e.g.} an insulating barrier) between two bulk superconductors, 
even in the absence of a voltage difference.\cite{josephson62}  This remarkable prediction by Josephson was already observed at the time by Giaever,\cite{giaever61} but was misinterpreted
as a ``metallic short.''\cite{giaever74} Both were awarded the
Nobel prize in 1973 for their discoveries.


Since its discovery, the Josephson effect has had a major impact on a broad spectrum of technologies. The most sensitive magnetic flux 
and electromagnetic radiation detectors are SQUIDs (Superconducting Quantum
Interference Devices), consisting of two Josephson junctions
connected in parallel. SQUIDs play a crucial role in many
condensed matter experiments and in radio-astronomy,
and recently in biomagnetic detectors to monitor brain activity. Josephson junctions were also a stepping stone in computer technology.  
In the 1970's, the burgeoning semiconductor revolution led IBM to abandon its effort to construct computers out of arrays of Josephson junctions. 
But today, circuits made of Josephson junctions are back in vogue, 
in the effort to construct a {\it quantum} computer.\cite{vion02, mcdermott05} 


Josephson's prediction was based on the fundamental principle of broken
symmetry.  A superconductor can be thought of in terms of a complex-valued wave-function
(or ``order parameter") $\psi(r)$ for Cooper pairs.
At the onset of superconductivity, Cooper
pairs undergo Bose condensation,
and the order parameter becomes nonzero with a  well-defined phase
$\varphi$: $\psi(r)=|\psi|\exp(i\varphi)$.  The symmetry of phase rotations is thus broken. 
When the phase $\varphi(r)$ varies in space, a bulk supercurrent proportional to its gradient
results. Similarly, a Josephson supercurrent flows between two superconductors 
with different phases. It
is proportional to the sine of the phase difference across the
weak link:
\begin{equation}
I= \frac{2 e J}{\hbar} \sin\left(\varphi_1-\varphi_2\right).
\label{eqn:josephson}
\end{equation}
Here, $J$ is the Josephson coupling energy, which characterizes the coupling strength between the superconductors.


The Josephson effect was initially measured between two {\it bulk} (\emph{i.e.}
three-dimensional) superconducting electrodes. Experimental
technique has since evolved dramatically. Present-day experiments
probe superconductivity in nanometer-scale samples, as well as in systems with reduced dimensionality. Particularly intriguing recent experiments measured the electrical resistance
across nanowire junctions between thin-film superconductors,\cite{bezryadin00,lau01} through a narrow constriction (about 20~nm wide) between two films as thin as 2.5 nm,\cite{chu04} and between a film and a superconducting scanning tunneling tip.\cite{naaman01}
In such nanoscale systems with reduced dimensionality,
the simple Josephson effect of
Eq.~(\ref{eqn:josephson}) must be revisited.


In large bulk superconductors, 
the phase of the order parameter $\varphi$ is a rigid,
essentially classical, variable. In  equilibrium, and in the absence of currents, $\varphi$ is locked to a single fixed value; the system exhibits
long-range order (LRO). 
This is true even in the presence of thermal fluctuations, provided the sample is
below the superconducting transition temperature.
The phase-rigidity of three-dimensional (3d)
superconductors prevents phase fluctuations that would suppress the
Josephson supercurrent, and Eq.~(\ref{eqn:josephson}) holds.  It is important to emphasize that these statements hold in the limit of large superconducting electrodes; otherwise, there is neither true LRO nor a true Josephson effect.

In thin-film superconductors, thermal fluctuations of the phase dramatically
alter this simple picture.
Specifically, phase correlations are no longer infinitely long-ranged; instead,
the correlator $\langle
e^{i\varphi(r_1)}e^{-i\varphi(r_2)}\rangle$ (where $\langle\dots\rangle$ denotes a thermal average)
decays as a power law with spatial separation.  This phenomenon is known as quasi-long-range order (QLRO), and occurs below the Berezinskii-Kosterlitz-Thouless
phase transition temperature $T_{{\rm BKT}}$.  Thermal phase fluctuations varying smoothly in space, present due to thermal excitation of the superconducting plasmon mode, are responsible for the power-law phase correlations.  Only at $T=0$, where these are frozen out, does LRO obtain.
Remarkably the
resistance of a two-dimensional (2d) film vanishes for $T<T_{{\rm BKT}}$ despite the absence of LRO.  Above $T_{{\rm BKT}}$, however, the superconducting state is 
disrupted by topological vortex defects, around
which $\varphi(r)$ winds by an integer multiple of $2 \pi$.
A nonzero density of mobile vortices scrambles the phase; only short-ranged
superconducting correlations survive, and the above correlator decays exponentially with spatial separation. In addition, the resistance of the film is nonzero in this regime.
Below $T_{{\rm BKT}}$, isolated vortices are expelled
from the film or tightly bound into vortex-antivortex pairs, and QLRO results.  Below, we consider the fate of the Josephson effect across a point contact separating two films below $T_{{\rm BKT}}$.

The question of the Josephson effect in systems with reduced dimensionality has
been addressed previously, but in the context of one-dimensional (1d) superconducting wires.
At any nonzero temperature, the phase correlations in 1d are short-ranged,
and one expects a non-vanishing resistance even in the absence of a weak link.\cite{langer67,mccumber70}
This can be understood in terms of thermally excited phase-slip events.
At zero temperature, a 1d superconductor exhibits QLRO, and the resistivity vanishes.  This holds provided $g > g_c$, where $g$ is the superfluid stiffness measured in appropriate units, and is proportional to the cross-section of the wire.  $g_c$ is on the order of unity and is determined by the nature of any Umklapp scattering that may be present.  The effects of a weak link on a 1d superconductor have been addressed theoretically; provided $g > 1$, it was found that
at zero temperature the weak link ``heals" itself, and the resistance across
the point contact vanishes (except in the case of a wire with only a {\it single} transverse
electron mode at the Fermi energy).  At small nonzero temperatures, 
the resistance across the point contact is predicted to vanish as a 
power law in temperature:\cite{kane92}
\begin{equation}
R \propto T^{2(g - 1)}.
\label{eqn:R1d}
\end{equation} 
The Josephson effect is obliterated at nonzero temperature, in dramatic contrast
to the weak link between 3d superconductors.  (It should be noted that, although ref.~2 explicitly dealt with a single-channel quantum wire, the effective field theory is identical to that for a many-channel superconducting wire, and the preceding statements follow immediately from the analysis there.)

What is the fate of the Josephson effect in the intermediate case, when the point contact is between two {\it films} rather than wires or bulk electrodes?
Surprisingly, despite several recent experiments on 2d films, very little theoretical attention has been paid to this problem (but see ref.~14).
In this Letter we determine the tunneling resistance across a point contact between two films 
in the geometry shown in Fig.~1.  As it turns out, a weak link between two films is \emph{almost  superconducting},
but a true Josephson effect is absent except at zero temperature.
Specifically, we find that the resistance $R(T)$ drops very rapidly upon cooling,
in a nearly activated fashion.
At low temperatures
\begin{equation}
\frac{R(T)}{R_Q} = \frac{t^2_v}{\sqrt{E_A(T)}} \frac{1}{(k_B T)^{3/2}}
\exp \Big[ -\frac{E_A(T)}{k_B T} \Big],
\label{RofT}
\end{equation}
where $t_v$ is an amplitude for quantum phase-slip processes,
discussed in more detail below.  
This formula is expected to be asymptotically exact at very low temperatures (see below).
Here $R_Q = h/ 4 e^2$ is the quantum of resistance, and $E_A(T)$ is a temperature-dependent activation energy:
\begin{equation}
\label{eqn:eaofT}
E_A(T) = c K_s  
\frac{1}{\ln(\hbar \omega_c/ 2 k_B T) + (2c K_s /  \pi^2 J)} .
\end{equation}
$R(T)$ is positive for all nonzero
  temperatures, but vanishes in an {\it activated} fashion as $T \to 0$, up to the
  logarithmic correction contained in the denominator of $E_A(T)$. $R(T)$ thus vanishes faster than any power law in the low-temperature limit. 
The scale of the activation energy is set by the superfluid stiffness in the 2d films,
$K_s = \hbar^2 n_s / m$, where $n_s$ is the density of Cooper pairs in
the film and $m$ is the pair mass.   
The dimensionless number $c$ is given by
\begin{equation}
c= \frac{\pi^2 \theta }{4 \alpha} ,
\end{equation}
where $\theta$ is the ``opening angle'' shown in
Fig.~1, and $\alpha$ is a parameter characterizing the
range of the interactions.  Specifically, 
$\alpha = 2$ for Coulomb interactions and
$\alpha = 1$ for screened interactions, as can be obtained
in the presence of a superconducting ground plane (see below). 
Remarkably, the main effect of the Josephson coupling
is to set a crossover temperature
$T_J \sim (\hbar \omega_c / 2 k_B) {\rm exp}(- 2c K_s/ \pi^2 J)$, where $\omega_c$ is a high energy cut-off discussed further below. 
At low temperatures, where the $\ln(1/T$) dominates
the denominator of $E_A(T)$, \emph{i.e.} $T \ll T_J$, the dependence on $J$ disappears altogether, 
and one recovers the {\it universal} result
\begin{equation}
E_A \sim 
c K_s/
{\ln (\hbar\omega_c / 2 k_B T)} .
\label{eqn:universal-limit}
\end{equation}
The important temperature scale in this limit is set by $K_s$, a property of the 2d films that can be measured independently\cite{fiory83} and is independent of the details of the contact.
At higher
temperatures, $R(T)$ is approximately given by a purely activated form (\emph{i.e.}~no logarithmic corrections), with the
barrier height set by the Josephson coupling:
$E_A \approx\frac{\pi^2}{2} J$.  

The resistance formula of Eq.~(\ref{RofT}) is exact at low temperature in the sense that \\
$\ln [R(T)] /  \ln (R_{{\rm measured}}) \to 1$ as $T \to 0$.  The
ratio of the resistances themselves does \emph{not} go to unity, as
there are additive corrections to $E_A(T)$
proportional to $\ln^{-2}(1/T)$.  These corrections are
contained in the integrals leading to Eq.~(\ref{RofT}) (supplementary information), so one can
do better by evaluating them exactly, which must be done numerically. 

Our result for $R(T)$ fulfills some basic physical requirements.  The zero-temperature
resistance vanishes, as it should in the presence of the true LRO
that obtains in the 2d films.
Furthermore, the resistance vanishes faster than any power law  as $T$ approaches zero, as is reasonable upon comparison with the 1d and 3d cases.  Finally, the activation energy
increases monotonically with increasing Josephson
coupling.



Before addressing the experimental implications of our result, we briefly discuss its derivation
and the underlying physics.  The thin-film superconducting electrodes and the point contact 
are modeled by a quantum phase Hamiltonian,
which focuses on the quantum and thermal fluctuations of $\varphi(r)$.  
This is legitimate at temperatures well below the
quasiparticle gap scale and $T_{{\rm BKT}}$.  
The Hamiltonian for a single thin-film electrode is, 
\begin{equation}
{\cal H}_{{\rm film}} = \frac{K_s}{2} \int d^2 r ( \nabla \varphi )^2 +
\frac{1}{2} \int d^2 r\, d^2 r'\, n(r) n(r') V(r - r'),
\label{eqn:film-hamiltonian}
\end{equation}
where $n(r)$ is the
fluctuating Cooper-pair density, canonically-conjugate to $\varphi(r)$.
The spatial integrations range
over the area of the film.  The first term encodes the energy cost for phase gradients; this is
essentially the kinetic energy of the superflow.  The
second term is a density-density interaction; we
consider both Coulomb [$V(r) = (2 e)^2/|r|$] and screened [$V(r) = (
\hbar^2 v^2_s / K_s ) \delta(r - r') $] interactions.  This Hamiltonian describes
the quantum dynamics of the plasmon mode of the superconductor.
In the case of screened interactions, there is a linearly dispersing acoustic plasmon ($\omega = v_s k$), whereas for
Coulomb interactions the plasmon has the dispersion $\hbar \omega = \sqrt{8 \pi e^2 K_s k}$.  The high-energy cutoff $\omega_c$ appearing in Eq.~(\ref{RofT}) is given in terms of this dispersion: $\omega_c = \omega(k = 2\pi/\xi)$, where $\xi$ is a short-distance cutoff on the order of the superconducting coherence length.

The second element in our system is the weak link, 
which we shall  refer to as a point contact. We model it as a
point-like Josephson coupling between the two
electrodes, and set the coordinate origin for both films (\emph{i.e.} $r=0$) at the contact:
\begin{equation}
{\cal H}_{{\rm contact}} = - J\, {\rm cos} (\varphi_2(0) - \varphi_1(0)).
\label{eqn:contact-hamiltonian}
\end{equation}
Here $\varphi_{1,2}(r)$ are the phase fields in the 
two electrodes.  This term can be interpreted as a process hopping Cooper pairs across the contact, where $J$ is the hopping amplitude.  It should be noted that we have neglected inter-film Coulomb interactions.  Because any geometric, zero-frequency capacitance will diverge with the size of the films, Coulomb blockade effects will be unimportant except below an extremely low temperature.  Above this temperature, we do not expect inter-film interactions to modify the low-temperature resistance.


In the limit $J \ll K_s$ (\emph{i.e.}~a poor contact), we may calculate the current response to a small
voltage bias in terms of the hopping of Cooper pairs across the
contact.  Specifically, one can expand in powers of $J$ to calculate the
conductance $G$ across the point contact: $G(T) = a_2 J^2 + O(J^4)$.
Kim and Wen\cite{ybkim93} found that the coefficient $a_2(T)$ is {\it divergent}
for temperatures below $T^* = \theta K_s / 2 \alpha k_B$, and concluded that the conductance diverges (and hence the resistance vanishes) for $T < T^*$.  However, the physical meaning of this result is unclear; in fact, we shall show that it indicates 
not zero resistance but rather a breakdown of perturbation theory.
More
information, and an independent test of this perturbative result, can be
obtained via an expansion in the opposite limit: that of a good
contact. Indeed, the above result already suggests a strong tendency towards
this limit, and justifies the approach taken below.

The limit of a good contact is readily accessed through a dual picture.
Here, we begin by assuming perfect phase-coherence across the contact.  This coherence is then weakened
by quantum phase-slip events, which occur when a
vortex tunnels across the film near the contact, in the direction
transverse to the current flow (as illustrated in
Fig.~1).  A phase slip  makes the relative phase of
the electrodes wind by $2\pi$; this $2\pi$ twist can either heal
locally by a phase slip in the opposite direction, or propagate
outward into the electrodes.  The latter process will register as a
voltage spike across the system, due to the Josephson relation
\begin{equation}
\Delta V = \frac{1}{2 e} \frac{d}{d t} \Delta \varphi .
\end{equation}

\begin{figure}
\includegraphics{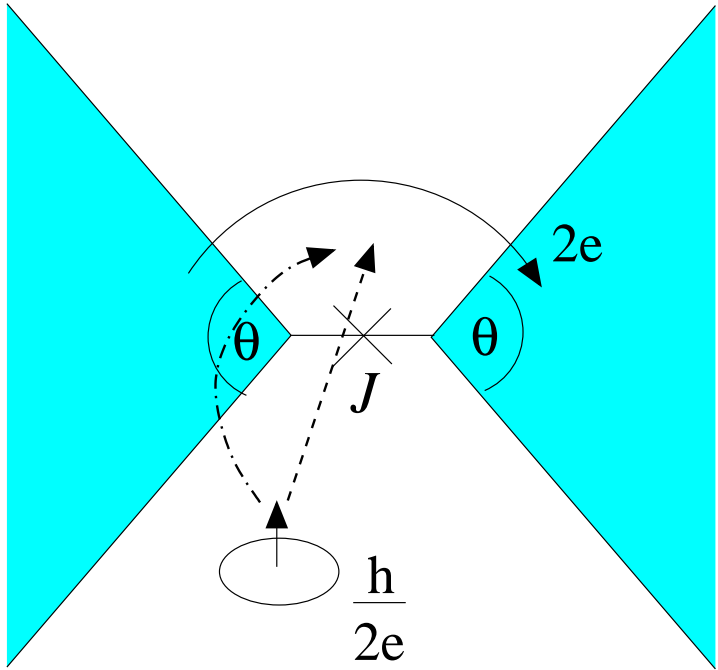}
\caption{Illustration of point-contact Josephson tunneling between two thin-film superconductors (shaded regions),  in the wedge geometry characterized by the angle
  $\theta$. This geometry directly models the experimental setup of ref.~6.  Charge transport across the contact can be viewed most directly in terms of the hopping of Cooper pairs (solid arrow) with amplitude given by the Josephson coupling energy $J$.  Alternatively, one can consider phase-slip events, in which the phase
difference between the two films winds by $2\pi$. These events occur
when a vortex hops across one of the films near the junction, transverse to the current flow (dash-dot line); this produces a momentary voltage spike between the films.  Because vortices can be ignored within the films at low temperature, all such events can be lumped together into a single effective process where a vortex tunnels directly across the contact (dashed line) with amplitude $t_v$.
As we argue in the text, it is essential to adopt the phase-slip point of view to understand the low-temperature transport.}
\label{fig:setup}
\end{figure}

In this dual picture the resistance can be obtained through a
perturbation expansion in $t_v$, which is the amplitude for a phase
slip to occur, or, equivalently, for a vortex to hop through the
contact.  Formally, we integrate out the degrees of freedom in the
films to obtain an action for the phase difference $\phi = \varphi_1(0)
- \varphi_2(0)$ across the contact. The
dual action for the phase slips can then be obtained by a Villain
transformation.\cite{villain75}  The phase slips are instanton events
where $\phi$ jumps by $\pm 2\pi$, and  we define $t_v$ as their fugacity.
The single vortex hopping process with amplitude $t_v$ should be viewed as an  encapsulation of the many different physical vortex-hopping processes in the vicinity of the contact (see Fig.~\ref{fig:setup}).
We note that $t_v$ depends only weakly on temperature (it goes to a constant at $T=0$).  It also has an implicit dependence on $J$, decreasing as $J$ increases.
It would be interesting to compute this dependence, 
but we do not attempt to do so here.

Following the above approach, 
we have calculated the voltage response to a vanishingly small current
across the point contact, obtaining the resistance
formula Eq.~(\ref{RofT}) at order $t^2_v$.   This formula encapsulates
the exact low-temperature dependence of the $t^2_v$ term, and is
quantitatively accurate over a reasonably large temperature range.  As an example, using the parameters for the system of MoGe films with ground plane discussed below, one can compare $R(T)$ to the exact resistance at order $t_v^2$ (obtained numerically).  The logarithms of these quantities agree within $15\%$
for $T \leq 4 {\rm K}$.
  Furthermore, contributions
of higher order in $t_v$ involve larger effective activation barriers
and are unimportant at low temperatures.  This is to be expected, as
these terms correspond to processes involving more phase-slip events,
and the phase slips are strongly suppressed by the QLRO in the films.

The resistance formula obtained from the dual calculation is expected to be exact at low temperatures,
even in the case of a poor contact.  Physically, the expansion in
phase-slip events is valid in the limit of small resistance.  As $R(T=0)$ vanishes due to the presence of LRO in the films, the phase-slip expansion
is always correct at low temperatures.  Reasoning along the same lines,
our result should be a good approximation as long as the argument of
the exponential in Eq.~(\ref{RofT}) is large and negative.

The most interesting regime is the universal low-temperature limit, \emph{i.e.} Eq.~(\ref{eqn:universal-limit}).  This regime will be more easily accessible when $J/K_s$ is rather large; this could be achieved by fabricating the point contact as a short and wide constriction.  

The most important quantity determining the universal low-temperature resistance is the
superfluid stiffness $K_s$.  This can be obtained directly for thin
films by an inductance measurement.\cite{fiory83}  In fact, one could
make a direct check of our theory by varying $K_s$ \emph{in situ} with an
in-plane magnetic field. The other relevant
parameters are the Josephson coupling $J$ and the cutoff
frequency $\omega_c$, which should be considered fitting parameters.  In many
cases it should be possible to estimate $J$ from $K_s$
and the geometry of the contact (see below).  An estimate of
$\omega_c$ is given in terms of the coherence length in the films, as discussed
above.


Our result should be accessible to experimental tests in a variety of
different systems.  Here we consider the
specific case, which was already realized by Chu \emph{et al.}, of a narrow constriction between two MoGe films.\cite{chu04}  We imagine a modification of their setup, in which a superconducting ground plane with stiffness $K^g_s$ is added a distance $d$ below the 2d films, which we take to be $100\,{\rm nm}$.  We consider $K^g_s \gg K^{\vphantom g}_s$, as appropriate for a thick ground plane, to screen out the Coulomb interactions as much as possible.  This setup leads to an acoustic plasmon propagating in the films with velocity $v_s = \frac{1}{\hbar} \sqrt{16\pi e^2 K_s d}$.  There is also a plasmon with $\omega \propto \sqrt{k}$ dispersion, which propagates within the ground plane and does not contribute to the tunneling transport.  
The constrictions in the experiments of ref.~6 had width $w \approx 20\,
{\rm nm}$ and length $\ell \approx 100\, {\rm nm}$.  The films
themselves exhibit a superconducting transition at $T_{{\rm BKT}}
\approx 5\, {\rm K}$, and have a coherence length $\xi \approx 7\, {\rm nm}$.
We consider $\theta = \pi$ and would have $\alpha = 1$ with the screening ground plane in place.
For illustrative purposes, we
assume  $K_s/ k_B \approx 10\, {\rm K}$ for $T \ll T_{{\rm BKT}}$, 
and then $\hbar \omega_c/ k_B \approx 2\pi \hbar v_s / k_B d \approx 2000  {\rm K}$.  (In the presence of the ground plane, the short-distance cutoff in the films is set by $d$ rather than $\xi$.) We estimate $J / k_B \approx
K_s w/ \ell k_B \approx 2\, {\rm K}$.  For these parameters, neglecting the temperature-dependence of $K_s$, the
argument of the exponential in $R(T)$ is $-1$ for $T \approx 5 {\rm
K}$, so we conclude that our formula should be good for $T \lesssim 5 {\rm
K}$.

One of the most important considerations for experiments is that
the superconducting films be in the thermodynamic limit.  
For a film of linear dimension $L$, we can define a crossover temperature scale $k_B T_L = \hbar\, \omega(k = 2\pi /L)$, which vanishes as
the linear size $L$ of the films diverges.  When $T \gg T_L$,
thermal processes prevent quanta of the plasmon mode from
traveling coherently between the film edges.  This thermal decoherence washes out the influence of the environment (and other finite-size effects) on the system of films and point contact.
With a superconducting ground plane as above, one has
$k_B T_L =   2\pi \hbar v_s /L$.  For 
$L=1 {\rm cm}$ we find $T_L \approx 2 \times 10^{-2}\, {\rm K}$.  
This leaves a broad temperature range over which the resistance formula Eq.~(\ref{RofT}) should apply, and also points out that it is important to have rather large films to lower $T_L$.  It is difficult to achieve such a broad temperature range without the ground plane, as $T_L$ is only proportional to $L^{-1/2}$ in that case; for the parameters as above, but with no ground plane, $T_L \approx 1\, {\rm K}$.  However, surrounding the film by a medium with rather large dielectric constant $\epsilon$ would lead to a modest improvement, because $T_L \propto \epsilon^{-1/2}$.

It is interesting to point out that the system studied here cannot be understood in terms of the phenomenological RCSJ (resistively and capacitively shunted junction) model, which is one of the standard theoretical approaches to Josephson junction physics.\cite{tinkham-book, schoen90}  In that model, one imagines that the junction is shunted by a parallel  resistor and capacitor.  One then writes down a Hamiltonian in terms of only $\phi$, the phase difference across the junction, and $N$, a canonically conjugate integer-valued Cooper pair density.  This takes the form:
\begin{equation}
{\cal H}_{{\rm RCSJ}} = \frac{1}{2 C} N^2 - J {\rm cos} (\phi) ,
\end{equation}
where $C$ is the capacitance.  The resistor is a phenomenological model for the dissipation in the system, which is assumed to be of the Caldeira-Leggett (CL) form\cite{caldeira81} appropriate for dissipation due to the plasmons of a 1d superconductor,\cite{kane92} or the similar Ambegaokar-Eckern-Sch\"{o}n (AES) form\cite{ambegaokar82} that models dissipation due to a Fermi surface of gapless quasiparticles or electrons.  The RCSJ model does not apply, however, for point contacts between large 2d or 3d superconducting leads with gapped quasiparticles.  In these cases, the capacitance diverges as the size of the leads is taken to infinity.  The charging energy vanishes in this limit, and Coulomb blockade effects disappear above a temperature scale proportional to $1/L$ for both 2d and 3d cases with Coulomb interactions.  As for the resistor, in these systems there is either dissipation due to the superconducting plasmons in 2d, or a pinning effect due to LRO in 3d (supplementary information).  These effects cannot be modeled by dissipation of the CL or AES form, as they are much stronger at low frequency.  Roughly speaking, the effect of the 2d or 3d electrodes is much closer to connecting the junction to a pair of ground planes, rather than to a resistor.  Indeed, even if one adds a physical shunt resistor to the system, at low frequencies it will be shorted out by the ground planes.

Finally, we expect that our
result will describe transport in a variety of systems  -- our model
only requires that the low-energy excitations in the films can be
described by some bosonic degrees of freedom governed by an XY model in
its QLRO phase.  This physics is most readily accessible in 2d superconducting films, and we hope our work will stimulate more experiments on these intriguing systems.

\begin{table}
\begin{center}
\centerline{\begin{tabular}{c | c | c}
Spatial dimensionality & Resistivity of individual electrodes & Point contact tunneling resistance \\
\hline
$d = 3$ &  Zero & Zero \\
\hline
$d=2$ & Zero & \emph{Nearly-activated} \\
\hline
$d=1$ & Power-Law & Power-Law
\end{tabular}}
\caption{\label{tab:resistance} {\bf Comparison of the electrode resistivity, and tunneling resistance across a point contact, for superconductors of varying spatial dimensionality.} The entries of the table describe the behavior at low temperature.  The main result of our work, which completes this table, is the nearly-activated behavior of the resistance across a point contact between two 2d superconductors.}
\end{center}
\end{table}

\bibliography{points-short}

\begin{thebibliography}{10}
\expandafter\ifx\csname url\endcsname\relax
  \def\url#1{\texttt{#1}}\fi
\expandafter\ifx\csname urlprefix\endcsname\relax\def\urlprefix{URL }\fi
\providecommand{\bibinfo}[2]{#2}
\providecommand{\eprint}[2][]{\url{#2}}

\bibitem{josephson62}
\bibinfo{author}{Josephson, B.~D.}
\newblock \bibinfo{title}{Possible new effects in superconductive tunnelling}.
\newblock \emph{\bibinfo{journal}{Phys. Lett.}} \textbf{\bibinfo{volume}{1}},
  \bibinfo{pages}{251--253} (\bibinfo{year}{1962}).

\bibitem{kane92}
\bibinfo{author}{Kane, C.~L.} \& \bibinfo{author}{Fisher, M. P.~A.}
\newblock \bibinfo{title}{Transport in a one-channel luttinger liquid}.
\newblock \emph{\bibinfo{journal}{Phys. Rev. Lett.}}
  \textbf{\bibinfo{volume}{68}}, \bibinfo{pages}{1220--1223}
  (\bibinfo{year}{1992}).

\bibitem{bezryadin00}
\bibinfo{author}{Bezryadin, A.}, \bibinfo{author}{Lau, C.~N.} \&
  \bibinfo{author}{Tinkham, M.}
\newblock \bibinfo{title}{Quantum suppression of superconducitvity in ultrathin
  nanowires}.
\newblock \emph{\bibinfo{journal}{Nature}} \textbf{\bibinfo{volume}{404}},
  \bibinfo{pages}{971--974} (\bibinfo{year}{2000}).

\bibitem{lau01}
\bibinfo{author}{Lau, C.~N.}, \bibinfo{author}{Markovic, N.},
  \bibinfo{author}{Bockrath, M.}, \bibinfo{author}{Bezryadin, A.} \&
  \bibinfo{author}{Tinkham, M.}
\newblock \bibinfo{title}{Quantum phase slips in superconducting nanowires}.
\newblock \emph{\bibinfo{journal}{Phys. Rev. Lett.}}
  \textbf{\bibinfo{volume}{87}}, \bibinfo{pages}{217003}
  (\bibinfo{year}{2001}).

\bibitem{naaman01}
\bibinfo{author}{Naaman, O.}, \bibinfo{author}{Teizer, W.} \&
  \bibinfo{author}{Dynes, R.~C.}
\newblock \bibinfo{title}{Fluctuation dominated Josephson tunneling with a
  scanning tunneling microscope}.
\newblock \emph{\bibinfo{journal}{Phys. Rev. Lett.}}
  \textbf{\bibinfo{volume}{87}}, \bibinfo{pages}{097004}
  (\bibinfo{year}{2001}).

\bibitem{chu04}
\bibinfo{author}{Chu, S.~L.}, \bibinfo{author}{Bollinger, A.~T.} \&
  \bibinfo{author}{Bezryadin, A.}
\newblock \bibinfo{title}{Phase slips in superconducting films with
  constrictions}.
\newblock \emph{\bibinfo{journal}{Phys. Rev. B}} \textbf{\bibinfo{volume}{70}},
  \bibinfo{pages}{214506} (\bibinfo{year}{2004}).

\bibitem{bcs}
\bibinfo{author}{Bardeen, J.}, \bibinfo{author}{Cooper, L.~N.} \&
  \bibinfo{author}{Schrieffer, J.~R.}
\newblock \bibinfo{title}{Theory of superconductivity}.
\newblock \emph{\bibinfo{journal}{Phys. Rev.}} \textbf{\bibinfo{volume}{108}},
  \bibinfo{pages}{1175--1204} (\bibinfo{year}{1957}).

\bibitem{giaever61}
\bibinfo{author}{Giaever, I.} \& \bibinfo{author}{Megerle, K.}
\newblock \bibinfo{title}{Study of superconductors by electron tunneling}.
\newblock \emph{\bibinfo{journal}{Phys. Rev.}} \textbf{\bibinfo{volume}{122}},
  \bibinfo{pages}{1101--1111} (\bibinfo{year}{1961}).

\bibitem{giaever74}
\bibinfo{author}{Giaever, I.}
\newblock \bibinfo{title}{Electron tunneling and superconductivity}.
\newblock \emph{\bibinfo{journal}{Rev. Mod. Phys.}}
  \textbf{\bibinfo{volume}{46}}, \bibinfo{pages}{245--250}
  (\bibinfo{year}{1974}).

\bibitem{vion02}
\bibinfo{author}{Vion, D.} \emph{et~al.}
\newblock \bibinfo{title}{Manipulating the quantum state of an electrical
  circuit}.
\newblock \emph{\bibinfo{journal}{Science}} \textbf{\bibinfo{volume}{296}},
  \bibinfo{pages}{886--889} (\bibinfo{year}{2002}).

\bibitem{mcdermott05}
\bibinfo{author}{McDermott, R.} \emph{et~al.}
\newblock \bibinfo{title}{Simultaneous state measurement of coupled Josephson
  phase qubits}.
\newblock \emph{\bibinfo{journal}{Science}} \textbf{\bibinfo{volume}{307}},
  \bibinfo{pages}{1299--1302} (\bibinfo{year}{2005}).

\bibitem{langer67}
\bibinfo{author}{Langer, J.} \& \bibinfo{author}{Ambegaokar, V.}
\newblock \bibinfo{title}{Intrinsic resistive transition in narrow
  superconducting channels}.
\newblock \emph{\bibinfo{journal}{Phys. Rev.}} \textbf{\bibinfo{volume}{164}},
  \bibinfo{pages}{498--510} (\bibinfo{year}{1967}).

\bibitem{mccumber70}
\bibinfo{author}{McCumber, D.} \& \bibinfo{author}{Halperin, B.~I.}
\newblock \bibinfo{title}{Time scale of intrinsic resistive fluctuations in
  thin superconducting wires}.
\newblock \emph{\bibinfo{journal}{Phys. Rev. B}} \textbf{\bibinfo{volume}{1}},
  \bibinfo{pages}{1054--1070} (\bibinfo{year}{1970}).

\bibitem{fiory83}
\bibinfo{author}{Fiory, A.~T.}, \bibinfo{author}{Hebard, A.~F.} \&
  \bibinfo{author}{Glaberson, W.~I.}
\newblock \bibinfo{title}{Superconducting phase transitions in
  indium/indium-oxide thin-film composites}.
\newblock \emph{\bibinfo{journal}{Phys. Rev. B}} \textbf{\bibinfo{volume}{28}},
  \bibinfo{pages}{5075--5087} (\bibinfo{year}{1983}).
\newblock \bibinfo{note}{And references therein}.

\bibitem{ybkim93}
\bibinfo{author}{Kim, Y.~B.} \& \bibinfo{author}{Wen, X.-G.}
\newblock \bibinfo{title}{Effects of collective modes on pair tunneling into
  superconductors}.
\newblock \emph{\bibinfo{journal}{Phys. Rev. B}} \textbf{\bibinfo{volume}{48}},
  \bibinfo{pages}{6319--6329} (\bibinfo{year}{1993}).

\bibitem{villain75}
\bibinfo{author}{Villain, J.}
\newblock \bibinfo{title}{Theory of one- and two-dimensional magnets with an
  easy magnetization plane. ii. the planar, classical, two-dimensional magnet}.
\newblock \emph{\bibinfo{journal}{J. Phys. (Paris)}}
  \textbf{\bibinfo{volume}{36}}, \bibinfo{pages}{581--590}
  (\bibinfo{year}{1975}).

\bibitem{tinkham-book}
\bibinfo{author}{Tinkham, M.}
\newblock \emph{\bibinfo{title}{Introduction to Superconductivity}}
  (\bibinfo{publisher}{McGraw-Hill: New York}, \bibinfo{year}{1996}),
  \bibinfo{edition}{2} edn.

\bibitem{schoen90}
\bibinfo{author}{Sch{\"o}n, G.} \& \bibinfo{author}{Zaikin, A.~D.}
\newblock \bibinfo{title}{Quantum coherent effects, phase transitions, and the
  dissipative dynamics of ultra small tunnel junctions}.
\newblock \emph{\bibinfo{journal}{Phys. Rep.}} \textbf{\bibinfo{volume}{198}},
  \bibinfo{pages}{238--412} (\bibinfo{year}{1990}).

\bibitem{caldeira81}
\bibinfo{author}{Caldeira, A.~O.} \& \bibinfo{author}{Leggett, A.~J.}
\newblock \bibinfo{title}{Influence of dissipation on quantum tunneling in
  macroscopic systems}.
\newblock \emph{\bibinfo{journal}{Phys. Rev. Lett.}}
  \textbf{\bibinfo{volume}{46}}, \bibinfo{pages}{211--214}
  (\bibinfo{year}{1981}).

\bibitem{ambegaokar82}
\bibinfo{author}{Ambegaokar, V.}, \bibinfo{author}{Eckern, U.} \&
  \bibinfo{author}{Sch{\"o}n, G.}
\newblock \bibinfo{title}{Quantum dynamics of tunneling between
  superconductors}.
\newblock \emph{\bibinfo{journal}{Phys. Rev. Lett.}}
  \textbf{\bibinfo{volume}{48}}, \bibinfo{pages}{1745--1748}
  (\bibinfo{year}{1982}).

\end{thebibliography}


\begin{addendum}
 \item We gratefully acknowledge discussions with L. Balents, A. Bezryadin, A. Paramekanti and X.-G. Wen.  This research is funded by
  the U.S. Department of Defense NDSEG program (M.H.), the U.S. National Science Foundation (G.R. and M.P.A.F.), and the U.S. Department of Energy, Division of  Material Sciences [through the Frederick Seitz Materials Research Laboratory at UIUC] (P.G.).
 \item[Supplementary Information]
 Here, we provide further information on the technical aspects of our results.  We shall discuss the action for the phase difference $\phi$ across the point contact, the duality transformation to the phase-slip action, and the derivation of the resistance formula.  In this supplementary information, we work in units where $\hbar = 1$ and $k_B = 1$.

The partition function for the coupled system of films and point contact can be written as the phase-space functional integral
\begin{equation}
Z = \int {\cal D}\varphi_i(r,\tau) {\cal D}n_i(r,\tau) \exp\big(-S_1 - S_2 - S_J),
\end{equation}
where
\begin{equation}
S_i = i \int_0^\beta d\tau \int d^2 r \, \partial_{\tau} \varphi_i(r, \tau) n_i(r,\tau) + \int_0^\beta d\tau \, {\cal H}^i_{{\rm film}}(\tau)
\end{equation}
is the action for film $i$ ($i = 1,2$), and
\begin{equation}
S_J = -J \int_0^\beta d\tau \cos\big(\varphi_2(0,\tau) - \varphi_1(0,\tau) \big)
\end{equation}
is the action for the point-contact Josephson coupling.  A product over space, time and film indices is implicit in the integration measure; that is,
\begin{equation}
{\cal D}\varphi_i(r, \tau) \equiv \prod_{i = 1,2} \prod_{r, \tau} d \varphi_i(r, \tau) \text{,}
\end{equation}
and similarly for ${\cal D}n_i(r, \tau)$.
The Hamiltonian ${\cal H}^i_{{\rm film}}$ is defined in Eq.~(\ref{eqn:film-hamiltonian}), and $\beta \equiv 1/T$.  The fields $\varphi_i(r,\tau)$ and $n_i(r,\tau)$ satisfy periodic boundary conditions in imaginary time, \emph{e.g.} 
$\varphi_i(r,\tau + \beta) = \varphi_i(r,\tau)$.  Furthermore, at the edges of each film, the phase satisfies the boundary condition $\nabla \varphi_i \cdot \boldsymbol{v} = 0$, where $\boldsymbol{v}$ is the normal vector to the edge.  This is simply the physical requirement that no supercurrent flow across the edge.

Because the films are governed by a quadratic action, standard techniques can be employed to integrate out all the fields except $\phi(\tau) \equiv \varphi_2(0,\tau) - \varphi_1(0,\tau)$.  In order to do this, it is first convenient to transform the fields to a basis in which $S_i$ is diagonal.  In the case of short-ranged interactions (\emph{i.e.} $V(r-r') \propto \delta(r - r')$) this is entirely straightforward; the kinetic energy term is diagonal in a basis of eigenfunctions of the Laplace operator $-\nabla^2$ with the above boundary condition, and in this case the density-density interaction is diagonal in any basis.  In the case of an infinite, translation-invariant plane [\emph{i.e.} $\theta = 2\pi$], Coulomb interactions can also be treated in an entirely straightforward manner.  In general,
Coulomb interactions are somewhat more difficult to deal with, because it is nontrivial to simultaneously diagonalize the kinetic and interaction terms.  To avoid this difficulty, we replace the Coulomb interaction with an effective potential, which is constructed to satisfy certain key properties that we believe are enough to leave the final result unaffected.\footnote{Consider an infinite, translation-invariant plane.  There, the Coulomb interaction is diagonal in momentum space, and has the Fourier transform  
$\tilde{V}(k) = 8 \pi e^2 / |k|$.  The Laplacian is diagonal in a plane-wave basis, and we can define its square root by writing $\sqrt{- \nabla^2} e^{i k \cdot r} = |k| e^{i k \cdot r}$.  Therefore we can write
$\tilde{V}(k) = 8 \pi e^2 (\sqrt{ - \nabla^2})^{-1}$.
The new potential $\bar{V}(r,r')$ is constructed so that this equation also holds, except that $\sqrt{-\nabla^2}$ is defined in terms of eigenfunctions of $-\nabla^2$ satisfying the $\nabla \varphi \cdot \boldsymbol{v} = 0$ boundary condition.  The resulting form depends separately on $r$ and $r'$, and not only on the distance $|r - r'|$; however, as the problem in the wedge geometry already lacks translation and rotation invariance, it should be harmless to sacrifice these in the form of the potential, as far as universal properties are concerned.  Furthermore, and most importantly, it can be shown that $\bar{V}$ has the same scaling behavior as the Coulomb potential: $\bar{V}(s r, s r') = \frac{1}{s} \bar{V}(r, r')$, where $s$ is a positive real number.}

At this point, a straightforward calculation yields an  effective partition function depending only on $\phi$:
\begin{equation}
Z_{{\rm eff}} = \int {\cal D}\phi(\tau) \exp\big(-S_0 - S_J) ,
\end{equation}
where
\begin{equation}
S_0 = \frac{2 c K_s}{\pi^2} \frac{1}{\beta} \sum_{\omega_n} \left[ \ln \frac{\omega_n^2 + \omega_c^2}{\omega_n^2} \right]^{-1} | \tilde{\phi}(\omega_n)|^2 .
\end{equation}
Here, $\omega_c$ is a high-frequency cutoff set by the short-distance cutoff in the films, $\omega_n \equiv 2\pi n / \beta$ ($n$ integer) are the usual Matsubara frequencies, and 
\begin{equation}
\tilde{\phi}(\omega_n) = \int_0^\beta d\tau e^{i \omega_n \tau} \phi(\tau) .
\end{equation}

It is interesting to compare this result to its 3d analog.  There one begins with the Hamiltonian for the bulk electrode
\begin{equation}
{\cal H}^{3d} = \frac{K^{3d}_s}{2} \int d^3 r (\nabla \varphi)^2
+ \frac{1}{2} \int d^3 r d^3 r' n(r) n(r') V(r - r') ,
\end{equation}
where $K^{3d}_s$ is the 3d superfluid stiffness in units of energy per unit length.  Proceeding as above, one finds
\begin{equation}
S_0^{3d} = \frac{K^{3d}_s}{\beta} \sum_{\omega_n} g(\omega_n) | \tilde{\phi}(\omega_n)|^2 ,
\end{equation}
where $g(\omega_n)$ goes to a nonuniversal constant as $\omega_n \to 0$.  This is a ``mass term'' for $\phi$, which localizes the phase and is expected to lead to a tunneling resistance of strictly zero (for superconducting leads in the thermodynamic limit).  This expectation is confirmed by a calculation analogous to that described below for the 2d case.

Furthermore, it is instructive to compare $S_0$ and $S_0^{3d}$ to the Caldeira-Leggett term that would be included in an RCSJ model.  This takes the form
\begin{equation}
S^{{\rm CL}}_0 \propto \sum_{\omega_n} | \omega_n | |\tilde{\phi}(\omega_n) |^2 ,
\end{equation}
which is clearly dominated by both $S_0$ and $S^{3d}_0$ at low frequency.  Therefore, even if it were appropriate to include a term of the form $S^{{\rm CL}}_0$ (it is not for the systems discussed here), it would have no effect on the low-temperature physics.

Now we return to the 2d system.  The partition function $Z_{{\rm eff}}$ is the starting point for the duality transformation that allows us to work directly in terms of phase slips.  The first step is to replace the cosine with a Villain function:
\begin{equation}
\exp \big(J \cos(\phi)\big) \to \sum_{\eta = -\infty}^{\infty} \exp \Big( - \frac{J}{2} (\phi - 2\pi \eta)^2 \Big)
\end{equation}
This is simply a different $2\pi$-periodic potential, the detailed form of which is not expected to affect universal properties.  Strictly speaking, when making this replacement one should also replace $J$ by an effective Josephson coupling (which is expected to be approximately equal to $J$).  However, because $J$ only enters the final result as a fitting parameter, it is not necessary to make this distinction here. 

We obtain the Villain partition function by making $\eta$ a periodic function of imaginary time ($\eta(\tau + \beta) = \eta(\tau)$):
\begin{equation}
Z_V = \int {\cal D}\phi \sum_{\eta(\tau) = -\infty}^{\infty} \exp \Big[ -S_0 - \frac{J}{2} \int_0^\beta d\tau \big(\phi(\tau) - 2\pi\eta(\tau) \big)^2 \Big].
\end{equation}
This can be specified more precisely by writing $\eta(\tau) = \int_0^{\tau} d\tau' \rho(\tau')$ and working in terms of $\rho(\tau)$, which can be interpreted as the imaginary-time density of phase slips.  (We can set $\eta(0) = 0$ without loss of generality.)  Due to the periodic boundary conditions, there are an equal number of ``positive'' ($\eta \to \eta + 1$) and ``negative'' ($\eta \to \eta - 1$) phase slips in each configuration summed over to form $Z_V$.  Letting $m$ be the number of positive phase slips in a given configuration, we can write
\begin{equation}
\rho(\tau) = \sum_{i = 1}^{m} \Big( \delta(\tau - \tau_i) - \delta(\tau - \bar{\tau}_i) \Big) ,
\end{equation}
where $\tau_i$ and $\bar{\tau}_i$ are the times at which positive and negative phase slips occur, respectively.  In $Z_V$, the sum over $\eta$ becomes a sum over $\rho(\tau)$, which can be defined precisely as
\begin{equation}
\sum_{\rho(\tau)} \equiv \sum_{m=0}^{\infty} \Big(\frac{t_v^m}{2^m m!} \Big)^2
\int d\tau_1 \dots d\tau_m \int d\bar{\tau}_1 \dots d\bar{\tau}_m .
\end{equation}
Note that the phase-slip fugacity $t_v$ has now explicitly appeared.  Upon transforming to Fourier space and integrating out $\phi$, we obtain a partition function only in terms of $\rho$:
\begin{equation}
Z_V = \sum_{\rho(\tau)} \exp \Bigg[
-\frac{1}{2 \beta} \sum_{\omega_n} \frac{J}{1 + \frac{\pi^2 J}{4 c K_s}  \ln \big( \frac{\omega_n^2 + \omega_c^2}{\omega_n^2} \big)}  \Big| \frac{2\pi \tilde{\rho}(\omega_n)}{\omega_n} \Big|^2 \Bigg]
\label{eqn:rho-action}
\end{equation}

The duality transformation is completed by observing that Eq.~(\ref{eqn:rho-action}) is equivalent to the following sine-Gordon theory:
\begin{equation}
Z_{{\rm dual}} = \int {\cal D}\theta(\tau) \exp \big(-S^{{\rm dual}}_0 + t_v \int_0^{\beta} d\tau \cos(2\pi\theta) \big) ,
\end{equation}
where
\begin{equation}
S^{{\rm dual}}_0 \equiv \frac{1}{2 \beta} \sum_{\omega_n} \Big[ \frac{1}{J} + \frac{\pi^2}{4 c K_s} 
\ln \Big( \frac{\omega_n^2 + \omega_c^2}{\omega_n^2} \Big) \Big] \omega_n^2 |\tilde{\theta}(\omega_n) |^2 .
\end{equation}
Expanding in powers of $t_v$ we obtain
\begin{equation}
Z_{{\rm dual}} = \sum_{n=0}^\infty \frac{t_v^n}{2^n n!} \sum_{\sigma_1 \cdots \sigma_n = \pm 1}
\int d\tau_1 \cdots d\tau_n \int {\cal D}\theta
\exp \Big[ - S^{{\rm dual}}_0[\theta] + 2\pi i \sum_{i=1}^n \sigma_i \theta(\tau_i) \Big] .
\end{equation}
At this point it is helpful to note that $\theta(\tau) \to (\theta(\tau) + {\rm constant})$ is a symmetry of $S^{{\rm dual}}_0$, which means that only ``neutral'' configurations with $\sum_i \sigma_i = 0$ will contribute.  This allows us to reorganize the above expression and write:
\begin{equation}
Z_{{\rm dual}} = \sum_{\rho(\tau)} \int {\cal D}\theta \exp \Big[ -S^{{\rm dual}}_0[\theta] + 2\pi i \int d\tau \rho(\tau) \theta(\tau) \Big]
\end{equation}
Upon integrating out $\theta$ we immediately obtain Eq.~(\ref{eqn:rho-action}).

These manipulations also show that the operator $\exp(2\pi i \theta(\tau))$ inserts a phase slip at time $\tau$.  As a result, the voltage across the contact (in a Hamiltonian formulation) is
\begin{equation}
\hat{V} = \frac{2 \pi t_v}{2 e} \sin(2 \pi \hat{\theta}) .
\end{equation}
The linear response resistance (at leading order in $t_v$) can be expressed in terms of a dual Kubo formula:
\begin{equation}
R(T) = - \lim_{\omega \to 0} {\rm Re} \Big[ \frac{i}{\omega} G^V_R (\omega) \Big],
\end{equation}
where
\begin{equation}
G^V_R(\omega) = -i \int_0^{\infty} dt \, e^{-i \omega t} \langle [\hat{V}(t), \hat{V}(0)] \rangle_{t_v = 0} .
\end{equation}
Note that the thermal average is evaluated in the limit $t_v = 0$, \emph{i.e.} in the quadratic theory governed by $S^{{\rm dual}}_0$.
This Green function can be obtained by analytic continuation from imaginary time, resulting in an integral expression for $R(T)$.  It is convenient to deform the contour in this integral, leading to the following simple form:
\begin{equation}
R(T) = \frac{\pi^2 t_v^2}{4 e^2 T} \int_{-\infty}^{\infty} dt\, e^{f(t)} ,
\end{equation}
where
\begin{equation}
f(t) = 16 c K_s \int_0^{\omega_c} \frac{dx}{x^2}
 \frac{\cos(x t) - \cosh(\frac{\beta x}{2})}{\sinh(\frac{\beta x}{2})}
 \frac{1}{\pi^2 + \big[\frac{4 c K_s}{\pi^2 J} + \ln \big( (\omega_c/x)^2 - 1 \big) \big]^2} .
 \end{equation}
 
 The function $f(t)$ has a maximum at $t=0$, and $f(0) \to -\infty$ as $T \to 0$.  Therefore, the saddle-point method can be used to evaluate $R(T)$ at low temperature; one finds
 \begin{equation}
 R(T) = \frac{\pi^2 t_v^2}{4 e^2 T} \sqrt{\frac{2\pi}{-f''(0)}} \exp\big( f(0) \big) .
 \end{equation}
 This formula is our main result, and its evaluation at low temperature leads to Eq.~(\ref{RofT}).  In particular, at asymptotically low temperature
 \begin{equation}
 \label{eqn:universal-formula}
 f(0) \sim -\frac{c K_s}{T \ln(\omega_c / 2 T)} .
 \end{equation}
 The full form of $E_A(T)$ given in Eq.~(\ref{eqn:eaofT}) is an approximation to $f(0)$ that reduces to the universal form at low temperature, and is valid over a larger range of temperatures.  For example, in the system of MoGe films in the presence of a superconducting ground plane discussed in the paper, $E_A(T)$ is within $15 \%$ of its exact value [obtained by a straightforward numerical evaluation of $f(0)$]  for $T \leq 4\, {\rm K}$.  On the other hand, the universal limit Eq.~(\ref{eqn:universal-formula}) only agrees within $50 \%$ in this temperature range.

\end{addendum}

\end{document}